# Physically Interpretable Diffractive Optical Networks for High-Dimensional Vortex Mode Sorting

Ruitao Wu,[†] Juncheng Fang,[†] Rui Pan, Rongyi Lin, Kaiyuan Li, Ting Lei,[*] Luping Du,[*] and Xiaocong Yuan[*]

*Nanophotonics Research Centre, Institute of Microscale Optoelectronics & State Key Laboratory of Radio Frequency Heterogeneous Integration, Shenzhen University, Shenzhen 518060, China.*

[*] Corresponding authors: leiting@szu.edu.cn, lpdu@szu.edu.cn, xcyuan@szu.edu.cn

[†] These authors contributed equally to this work.

**Abstract:**

Despite the significant progress achieved by diffractive optical networks in diverse computing tasks, such as mode multiplexing and demultiplexing, investigations into the physical meanings behind complex diffractive networks at the layer level have been quite limited. Here, for high-dimensional vortex mode sorting tasks, we show how various physical transformation rules for each layer within trained diffractive networks can be revealed under properly defined input/output mode relations. An intriguing physical transformation division phenomenon, associated with the saturated sorting performance of the system, has been observed with an increasing number of masks. In addition, we have also demonstrated the use of physical interpretation for efficiently designing parameter-varying networks with high performance. These physically interpretable optical networks resolve the contradiction between rigorous physical theorems and operationally vague network structures, paving the way for designing and understanding systems for various mode conversion tasks, and inspiring further interpretation of diffractive networks in advanced tasks and other network structures.

## Introduction

Optical diffractive networks (DNs) have been developed to perform computing at the speed of light in a power-efficient manner for diverse tasks or applications, including image classification[1], optical imaging[2], and quantum-related problems[3]. From a spatial mode manipulation perspective, the global functionality of a DN can be described by a linear matrix operator that performs input/output mode mapping[4-6]. Such operations are achieved through cascaded diffractive surfaces with spatially varying phase distributions, which are often referred to as multi-plane light conversion (MPLC) for the optical device community[7-9]. Perhaps the most interesting and well-known application of DNs/MPLCs is mode multiplexing/demultiplexing, since it directly relates to a crucial application, space-division optical communications[10, 11].

In general, the design of diffractive surfaces for operating modes can be classified into two approaches. In certain scenarios, analytical solutions for the phase distribution have been reported and demonstrated. This approach is typically applied to low-dimensional mode transformations with a small number of masks with clear mathematical operations, such as the sorting of orbital angular momentum (OAM) modes through dual-plane optical coordinate transformations[12, 13]. In most cases, especially when the dimensionality of the problem increases and no known solutions are found, it is necessary to apply iterative optimization and inverse design algorithms, such as wave-front matching (WFM)[14, 15] or the machine-learning framework[1], which have been proven to be both efficient and flexible.

Physically interpretable DNs dealing with mode sorting tasks are expected to be both intuitive and generalizable, which offers new insights for designing high-performance optical devices, such as spatial mode multiplexer/demultiplexer. However, the nature of iterative algorithms has inevitably resulted in "random" masks that bring challenges in understanding and interpreting the mathematical operations behind, especially given the large number of variables (pixels and layers), even though the networks perform only linear tasks[16, 17]. This issue is akin to the "black box" problem in deep learning (DL), as providing model interpretability for neural networks (NNs) is challenging (Fig. 1a)[18-20]. The difficulty of interpreting persists even in relatively simple mode conversion tasks, such as the sorting of Laguerre-Gaussian (LG) modes[14, 21]. With the rapid development of DNs and MPLCs, it becomes critical to tackle this issue. To our knowledge, while seemingly meaningful patterns have appeared in several related reports, clear physical interpretations at individual layers of reported models have never been discussed or provided[22-26].

Here, we present the first physically interpretable diffractive optical networks for LG mode sorting (Fig. 1b), where the functionality of each layer is unveiled. We show how various known physical transformation rules emerge after training the DN with specific input-output mode relations and dimensionalities. For high-dimensional mode sorting DNs, an intriguing physical transformation division phenomenon arises, accompanied by our observation of an optical analogy of the underfitting/overfitting effect in DL. Furthermore, we demonstrate how such a physical interpretation reduces the computational effort for designing high-performance, high-dimensional DNs in the presence of system variations. Our study not only provides new paths for understanding and designing high-performance DNs, but also has significant implications for research based on network configurations.

## Results

### Re-discovery of coordinate transformation rules within diffractive networks

For the mode sorting task that we discussed in this paper, the term physical interpretability would mean that the functionality for each mask within the DN should be intuitive and understandable[18]. For a mask with multiple levels of functionalities (e.g., simultaneously performs phase correction, beam shaping, or others), one should be able to identify and distinguish all of them. In other words, all physical processes within the trained DN should be easy to follow. We should emphasize that this is done for all layers, as previous studies have only focused on the global physical functionality. Besides, we will deal with the demultiplexing of the LG modes[21], which are described by two quantized indices, l, the topological charge, and p, the radial number (Supplementary Note 1). The terms "LG modes" and "high-dimensional vortex modes" are used interchangeably in this paper. The DNs are trained as sorters for optical modes according to these two indices. Theoretically, it can be considered as a conversion of Hilbert space from the mode index spectrum to spatial distribution[3]. We will show how conventional mode conversion rules emerge for these trained complex networks with different layers.

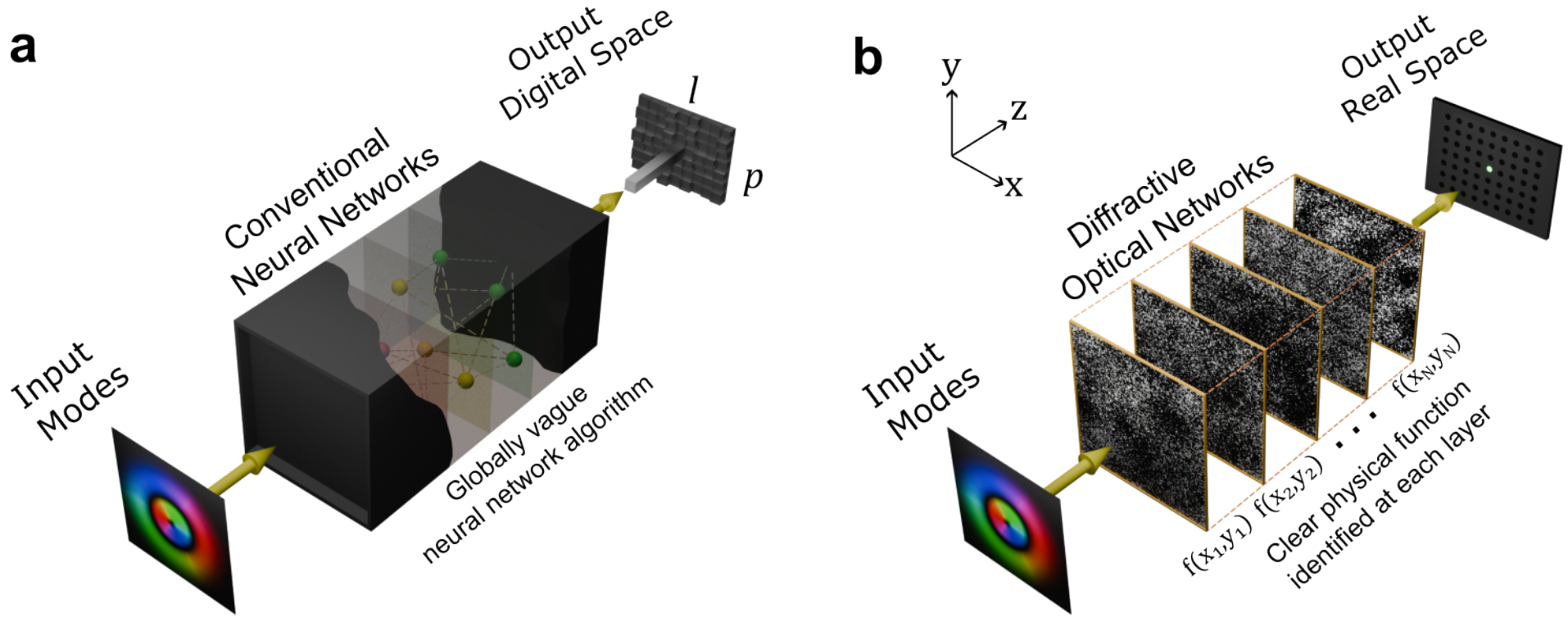

**Fig. 1. Physically interpretable diffractive optical networks for high-dimensional vortex mode sorting.** (a) Conventional digital neural network layout. The algorithm is typically considered a global "black box" since it gives the output in a way that is not physically intuitive and understandable. (b) A diffractive optical network is implemented in physical space to conduct the mode sorting task. In previous works, the physical meaning of this optical analogy neural network remains unclear, mainly due to the seemingly random phase masks and complex intra-layer connections. This work shows how to interpret such networks and unveil their physical meaning at the layer level.

In this section, we first consider a two-layer diffractive structure that can perform input LG mode sorting according to their orbital angular momentums (OAMs) (Fig. 2a)[27]. In other words, we will decrease the dimensionality of the problem by considering only the case of $p=0$. The case when $p\neq0$ will be discussed in the next section. The outputs are set to be Gaussian-like modes distributed along the x-axis. Our dataset consists of 11 LG modes with $l$ from -5 to 5, while $l$ = 4, 2, and -2 are randomly picked to be the test set, and the rest are used as the training set. More detailed information on the training process is provided in Materials and Methods. As we will show in the following, these two diffractive layers work synergistically to perform coordinate transformation as an afocal system and then perform the mode sorting task.

The trained masks are presented in Fig. 2b (top line), where an interesting continuous phase gradient can be observed within the region of interest (ROI), where most of the energy is distributed. The ROI region is enclosed by the red dotted lines for illustration purposes, which is determined by numerically propagating all the modes through the system and examining the intensity distribution on each mask. Later on, the trained network is physically implemented and tested on the MPLC system (Materials and Methods). The network performance is further characterized by the detection probability matrices (Fig. 2c), with an average of 77% (numerically) and 72% (experimentally).

What is interesting is that the shapes of ROIs (Fig. 2b) transform from a circle to a rectangle for two different masks, respectively. This reminds us about a critical functionality of diffractive surfaces, i.e., performing coordinate transformation[28, 29]. The task that the DN is trained for is the vortex mode sorting, a well-known problem in the field of optics. One of the elegant solutions is to apply the log-polar transformation, then use a lens to focus the transformed field in the output plane. This solution requires a minimum of two masks[12]. The first mask acts as a log-polar transformation unwrapper, and the second mask performs the phase correction and the focusing (Fig. 2a). It's noteworthy that transformation terms within these two masks strongly depend on each other since they synergically form an afocal system[29]. The exact parameters of both masks, however, can be retrieved from the trained network and calculated analytically (Supplementary Note 2). Therefore, masks of analytical solutions for the log-polar transformation method are evaluated and shown in Fig. 2b (bottom line). Surprisingly, the trained DN (top line) resembles the coordinate transformation theory very well within the ROIs. We further extracted the transformation term within these two mask pairs and verified that the functionality of our trained network performs the same as the aforementioned theoretical approach, as numerically evaluated and experimentally verified in Fig. 2d.

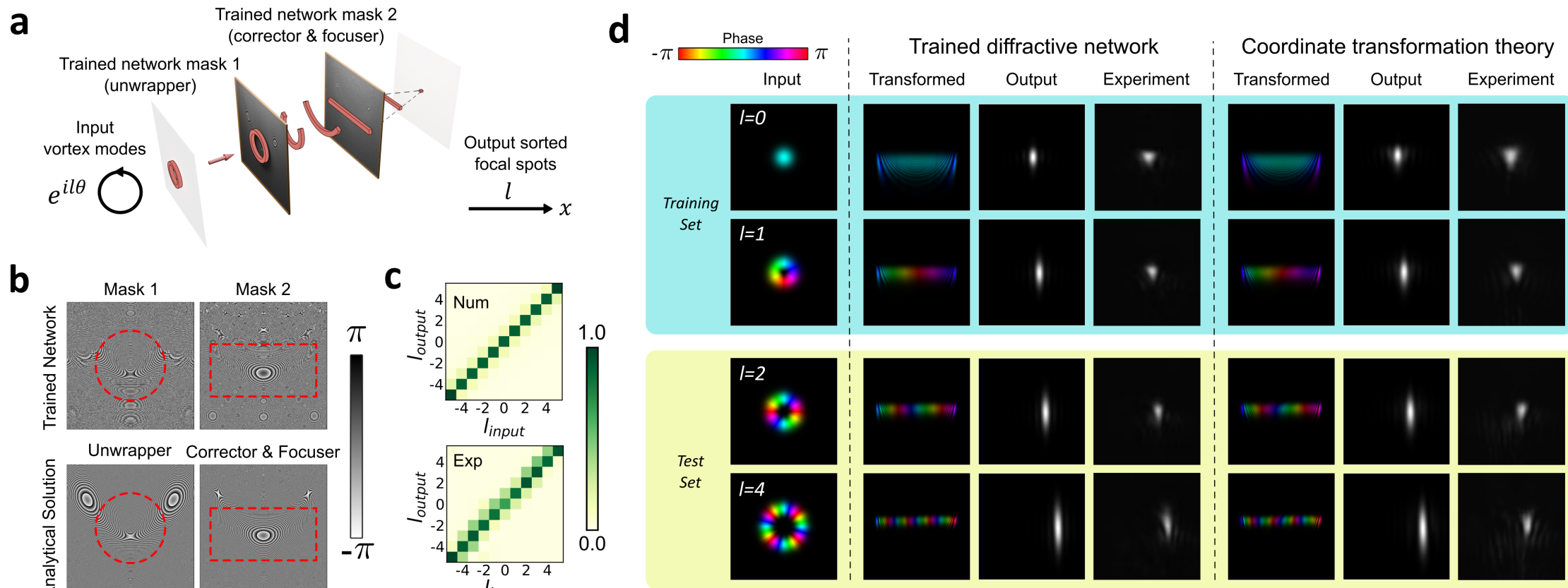

**Fig. 2. Physically interpretable diffractive network as a vortex mode sorter: re-discovery of the log-polar transformation approach.** (a) Sketch of the two-layer DN trained as a vortex mode sorter. (b) The trained mask pairs resemble the log-polar transformation, while one mask acts as the unwrapper and the other one serves as the corrector and the focuser. Red dotted lines indicate the region of interest, where optical power is focused in each layer. (c) The detection efficiency matrix for the trained network and the theoretical converter for mode sorting. (d) Verification of the transformation process for both the log-polar mode sorter and the trained network. The images show the phase of numerical model light fields (colorful images) and the intensity (grayscale images, both numerical and experimental data). We note that the intensity images are magnified by 4 times with respect to the field images.

Overall, we have shown how a DN can learn about the well-known "log-polar transformation and focusing" approach after training for a one-dimensional vortex mode sorting task. The physical functionalities with both layers are unveiled, while the resultant trained network matches the theoretical prediction well. This example proved that even for DN trained by iteratively optimized algorithms, it is possible to fully "transparentize" all the physical operations behind each layer. Other types of operations, such as circular-sector transformation and beam shaping, can be realized with such a system, once a proper relation between the input/output mode is defined, which is demonstrated in the following.

## Unveiling physics within diffractive networks for high-dimensional vortex mode sorting

Next, we show how different physical processes can be separated from each other for a trained network of high dimensionality. We move on to a mode converter that can simultaneously sort both indices of input LG modes into a 2D rectangle distribution. The dimensionality of the DN is increased by incorporating the radial index and assigning it to the y-axis, while the index l is distributed along the orthogonal x-axis. We would like to point out that, although the design of a seven-layer diffractive LG mode sorter has been experimentally demonstrated, the functionality of each diffractive layer within the diffractive network remains unclear[14]. Since the azimuthal index can be sorted through the aforementioned approach (requires two masks), one might naively expect that the sorting of radial factor p will require an independent physical transformation process (which requires at least two additional masks), as these two indices are independent from each other[30]. The masks for two independent transformation processes might be cascaded to reduce the number of masks[28]. Therefore, in this section, we will consider DNs with three diffractive layers. The training process is similar to the example in the previous section, with additional details given in Materials and Methods. Our dataset consists of 33 $LG_{l,p}$ modes with l from -5 to 5 and p from 0 to 2, while the cases $LG_{-2,2}$, $LG_{3,1}$, $LG_{-3,1}$, $LG_{4,0}$, $LG_{1,0}$ are randomly chosen to be included in the test set, and the rest are used as the training dataset. We note that the choice of test sets and training sets would only slightly modify the trained mask distribution and would not affect the physical meaning behind each layer (Fig. S1).

The physical transformation processes within the trained network are sketched in Fig. 3a. The functionalities of each mask can be initially evaluated by propagating the input fields to each sequential mask and calculating the diffractive fields as well as the ROI regions. Aside from the focuser in the last plane, two distinct and independent diffractive processes can be identified, with the detailed procedure discussed in Supplementary Note 3. The first diffractive process performs the circular-sector transformations[31], resulting in sector-shaped output fields carrying azimuthal phase gradients[12]. We note that such a transformation is slightly different from the reported works since it involves a translational term that performs center positioning. For the second transformation, the multiple sectors (depending on p values) will be reshaped into a relatively uniform circular distribution with a phase gradient along the y-axis, which can lead to repositioning after focusing. We would like to emphasize that there are no rigorous analytical solutions for such a process, and this is, in part, the reason why the radial factor p is challenging to sort using diffractive elements[32]. Nevertheless, the approximate mask solution for such a process might be estimated using the iterative phase matching algorithm discussed in Supplementary Note 3. Interestingly, the circular-sector transformation term always appears before the second reshaper term in all our trained outcomes. While other solutions (or layer-swapped DNs) are theoretically possible, the current solutions outperform and survive after multiple iterations of the algorithm. The resulting masks are shown in Fig. 3b (ROI region indicated by the red dotted lines). That being said, the physical functionalities of all three masks within such a network have been unveiled using our approach, as illustrated in Fig. 3b (and Fig. S3). The crosstalk matrices and the final output fields for the trained networks, which are numerically calculated and experimentally verified, are shown in Fig. 3c. Good agreement with the theoretical physical transformation is found. The transformation process is further illustrated in Fig. 3d, which verifies the functionality for each mask.

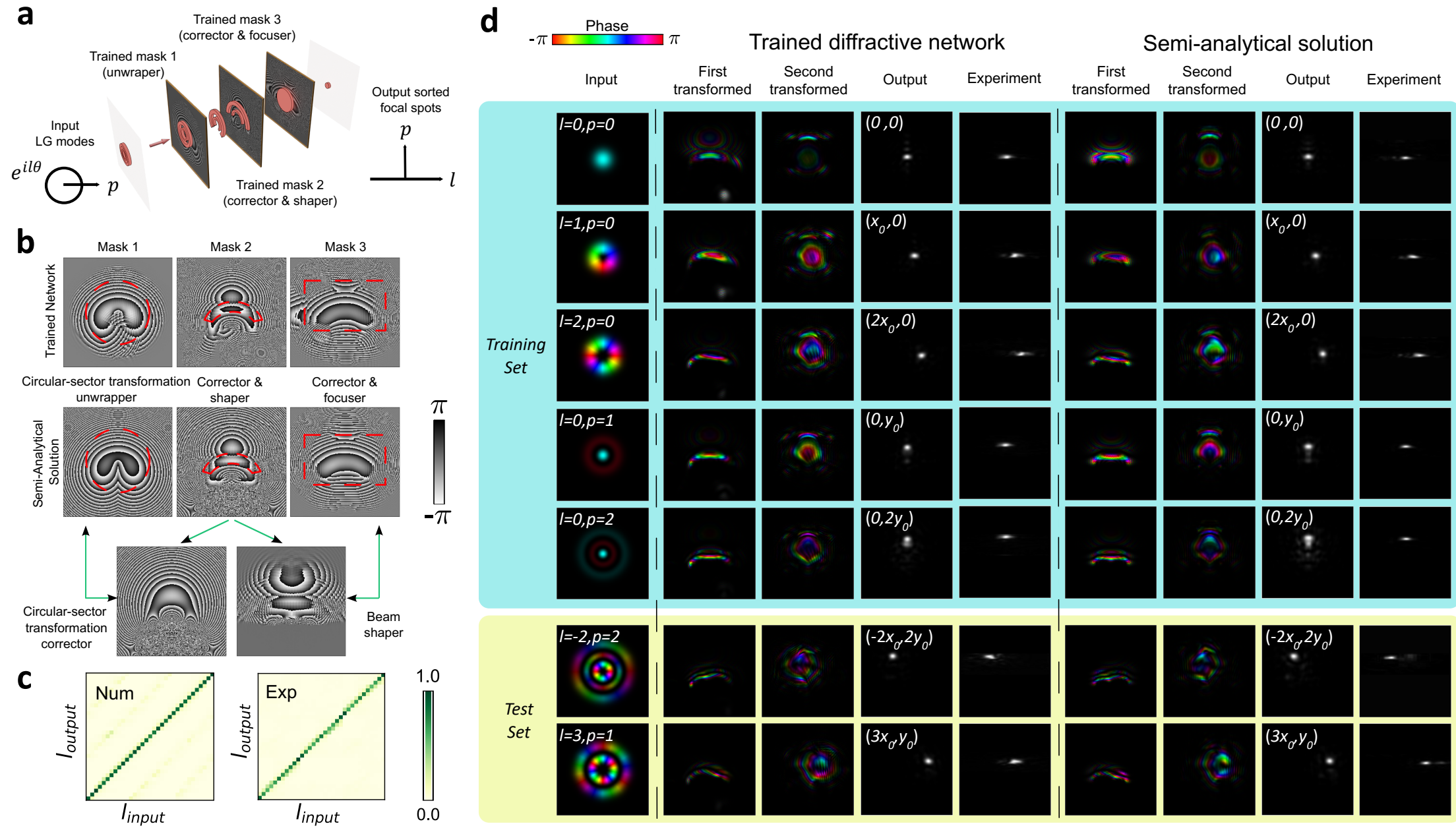

**Fig. 3. Physically interpretable triple-layer diffractive networks for high-dimensional vortex mode sorting: independence of multiple physical transformations.** (a) Sketch of the trained triple-layer DN for LG mode sorting, where the index l is sorted along the x-axis and the p value is distributed along the y-axis. (b) The LG mode sorting process in LG modes involves two transformations: 1. Circular-sector transformation. 2. A beam shaper turns the sector to a disk distribution with varying phase terms, as well as a radial lens that performs the focusing. Red dotted lines indicate the regime where optical power is focused. (c) The detection efficiency matrices for the trained network and semi-analytical converter as an LG mode sorter. (d) Verification of the transformation process for both the LG mode sorter and the trained network. The images show the phase of numerical model light fields (colorful images) and the intensity (grayscale images, both numerical and experimental data). We note that the intensity images are magnified by 4 times with respect to the field images.

We have demonstrated that the interpretation for such a high-dimensional mode conversion task with increasing the number of diffractive layers can be realized. Each layer has its own distinct yet identifiable physical operations. Additionally, what is interesting about this network, as compared to the previous literature[14], is that approximate transformation solutions can be found for both processes, which was unknown knowledge for the network before the training. The trained network extracts this information, making the sorting of the test set easy to understand. We envision that our physically interpretable diffractive optical network for LG mode sorting would be inspiring for designing other mode sorting systems, since it deepens our understanding during such a process.

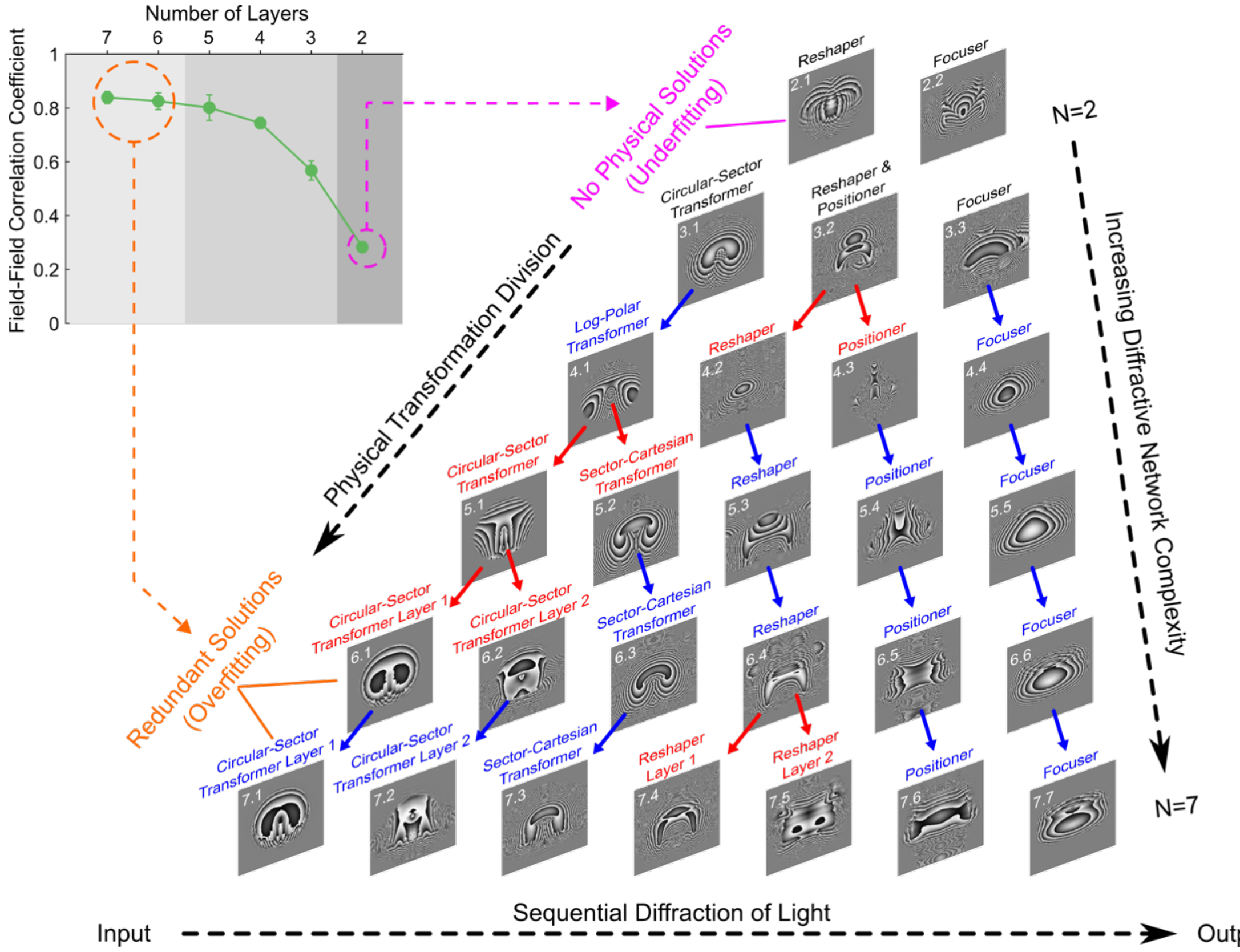

**Fig. 4. Physical transformation division phenomenon in multi-layer diffractive networks trained for LG mode classification.** When training a diffractive network for a specific mode classification task, with increasing network complexity (number of layers), the functionality of each diffractive mask can either be maintained (blue arrows) or split into two sequential layers (red arrows). The top left inset image shows the correlation coefficient between the designed and numerical output field, as a function of diffractive surface numbers. Error bars indicate the standard deviation of correlation coefficients over all the input modes. Such a physical transformation division effect can range from very few layers when there are no known physical solutions (underfitting, with an extremely low correlation coefficient), to multiple-layer cases in which redundant mask solutions appear (overfitting). In the intermediate regime (mask number from 3 to 5), clear physical transformation processes can be identified for each layer, while the system performance increases significantly. In the case of overfitting, the performance of the system does not benefit from the increasing number of masks since the correlation coefficient remains unchanged. We note that every mask in each system (except for the first mask) also performs a phase correction function, which is omitted for simplicity.

### Physical transformation division effect within high-dimensional diffractive networks

When designing such a network, the diffractive surface numbers in DNs have several constraints in practical aspects. Intuitively, small diffractive surfaces cannot convert complex modes with high purities[14]. On the other extreme, the alignment of a large number of masks can be technically challenging, not to mention the external environmental conditions like mechanical vibrations, or input scaling and rotations[33]. To our knowledge, while there are studies on the effect of surface numbers on the performance of DNs[34], the influence on the physical processes involved, particularly at each mask level, has never been investigated. Here, we investigate the evolution of all the physical processes when designing a DN for the LG mode sorting task.

We follow the discussion of our two-dimensional LG mode sorter and numerically extend it to the case of various diffractive mask numbers (denoted by N, from 2 to 7). Details regarding the training are similar to the example of three-layer systems, with additional details can be found in Materials and Methods. The trained DN and relevant results are summarized in Fig. 4. We would like to point out that, similar to the previous three-layer network example, every mask (except for the first) includes a term that compensates for the phase distortion induced by the previous mask. However, these phase correction functions are omitted for illustration purposes. Analysis of physical transformation in each mask is given in Supplementary Note 4.

An intriguing phenomenon appears when checking the evolution of physical processes for each mask, as shown in Fig. 4. We note that limited masks can not perform the sorting task with sufficient accuracy. Therefore, for a small number of layers, such as $N \leqslant 2$, no physical solutions are available for the LG mode conversion task. For N=3, we recover a similar system in the previous section, where a circular-sector transformation and a diffractor that can simultaneously reshape and reposition the beam are observed. When N is increased from 3 to 7, one can observe a "physical transformation division" process: the functionality of one diffractive mask for a DN of N layers will be performed by two consequent surfaces for DN of N+1 layers, as indicated by red arrows. For instance, when N increases from 4 to 5, the log-polar transformation is divided into two: a circular-to-sector transformation and a sector-to-cartesian transformation. Surprisingly, it has been demonstrated that a polar to Cartesian transformation can only be done with at least three surfaces[28] (that is why log-polar is used instead of polar for vortex mode sorting). Such knowledge is captured by the designed DNs. Starting from N=6, a "physically redundant" mask appears, as now a circular-sector transformation will require three masks instead of two (Mask 6.1 and 6.2). We attributed this interesting phenomenon to the physical transformation division effect. Interestingly, the observed "underfitting/overfitting" effect not only appears in the physical functionalities but also in the system performance. Specifically, the redundancy of these masks can be further verified by calculating the field-field correlation coefficient between the targeted field and the numerically evaluated output field, as now this parameter saturates after $N \geqslant 6$. We would like to emphasize that the field-field correlation coefficient is a strict quantity compared to the detection matrix since it requires not only the integrated intensity but also the local amplitude and phase distributions. This surprising correlation between the field-field correlation coefficient and redundant solutions has been identified, while the additional layers of the system do not improve the system performance.

Interestingly, the focusing phase within the system always appears as the last surface, as the exemplified network is trained as a sorter with performance evaluated by spatially distributed spots. This lensing effect in the last mask can also be found in other literature[1, 23]. This reminds us of another known effect in digital neural networks, the gradient vanishing effect[35], which sometimes leads to similar results as the overfitting effect, as the increasing number of neural layers has very little influence on the outputs. The difference is that the conventional gradient-vanishing effect typically appears in the first few network layers during the training, and typically leads to training failure[35]. In our case, the DN still exhibits good performance with the "gradient vanishing" layer playing a crucial role.

**Application of physical interpretation: extrapolation of pre-trained networks**

We devoted the last section to a promising application of physically interpretable DNs—designing new DNs with system variations by extrapolating previous trained DNs. Conventionally, a DN needs to be re-trained, regardless of what the system parameter is (and how much it changes). While it is possible to scale the pre-trained mask for a specific variable, other physical parameters will also be altered, which hinders its practicality. For instance, the extrapolation of a pre-trained DN for another wavelength will lead to different inter-plane distances[14]. Here, we demonstrated that, based on the knowledge of the mathematical operation behind, it is possible to extrapolate a pre-trained DN for a single varying parameter while maintaining satisfactory performance. This eliminates the need for retraining, which dramatically reduces the computation effort and time.

We would like to emphasize that such an extrapolation process does not require knowledge of analytical solutions for the transformation. It requires the pre-trained masks as well as the updated system parameters. Details of the extrapolation procedure are summarized in Supplementary Note 5. We first consider the aforementioned two-layer system designed for vortex beam sorting (p=0) as an example. Since the trained DN is found to be an optical system that performs both conformal mapping and the focusing process, we can apply the intrinsic property of coordinate transformation and extrapolate such a network. Numerical results on the extrapolation of previously trained two-layer DN to other wavelengths and interplane distances are summarized in Fig. 5a. We note that there is an intrinsic limited bandwidth of pre-trained masks for both wavelength (red curve) and distance (yellow curve). Nevertheless, it is possible to extrapolate the trained network and directly calculate the new mask without performing new training for a broad range of parameters. The performance of the new masks is identical to the pre-trained masks (indicated by solid dots).

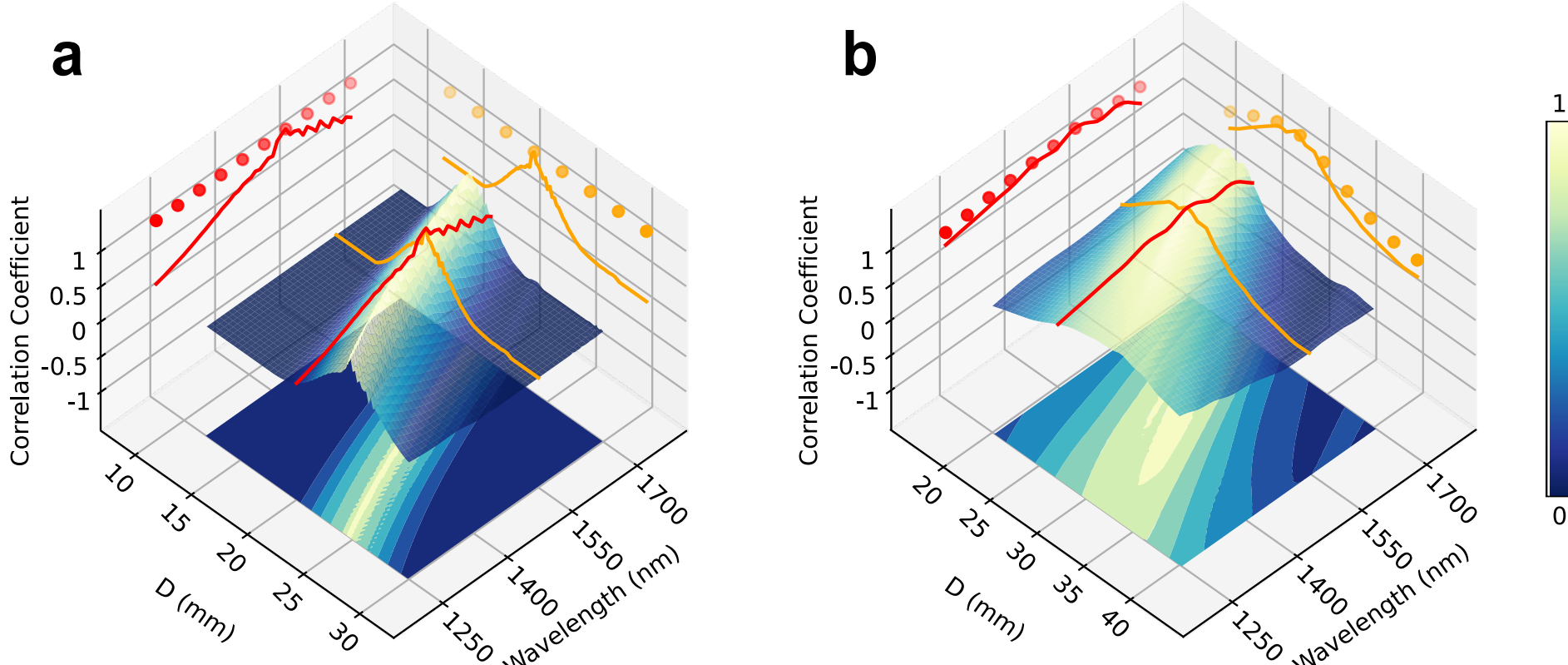

**Fig. 5. Extrapolation of physical interpretable diffractive networks under system variations.** (a) For the DN proposed for one-dimensional vortex mode sorting trained for single wavelength (1550 nm) and interplane distance (20 mm), the correlation coefficient of the designed sorted spots and output fields will severely degrade under system perturbation (see contour). The numerically extrapolated solution based on pre-trained masks can be numerically extended to other wavelengths (red dots) or distances (yellow dots) with unity correlation coefficients. (b) Results of numerical extrapolation performed for the three-layer two-dimensional LG mode sorting task. The pre-trained masks are designed for 1550 nm wavelength and 26.8 mm interplane distances. Similarly, the pre-trained network has an intrinsic bandwidth with respect to different parameters. The performance of the extrapolated solutions (wavelength: red dots; distance: yellow dots) is higher than the pre-trained masks. The correlation coefficient is smaller than the example in (a) due to the violation of the conformal mapping condition.

We note that the accuracy of this extrapolation relies on the conformal mapping condition, not the total number of layers within the system. This condition is met for sorting of azimuthal indices, including our two-layer DN example. However, the complex transformation involved in p-index sorting unavoidably violates this condition, as we discussed in previous sections. To validate this conclusion, we have performed extrapolation to the 3-layer network we discussed in Fig. 3, with the numerical results summarized in Fig. 5b. Similarly, such DN also has an intrinsic bandwidth for both wavelength (red curve) and distance (yellow curve). Since the conformal mapping conditions are not strictly satisfied in this case, the performance (characterized by the correlation coefficient of the numerically transmitted fields) of the extrapolated network is lower than unity. Nevertheless, the extrapolated network solutions (solid dots) have shown significant improvement in performance compared to the pre-trained masks.

We would like to emphasize again that this does not mean that all DNs with more than 2 layers can not be extrapolated with high performance. This issue arises when the conformal condition breaks. Besides, whether the conformal condition is violated or not can be known once the physical transformation terms are retrieved. Therefore, the designer is capable of determining whether such extrapolation can be performed or not before it is done. As long as this condition is met, our approach can also be applied to extrapolate DNs perturbed by other parameters, such as focal distance, input plane waist, etc. Furthermore, we noted that there is an interesting connection between our extrapolation approach and the transfer learning (a well-known approach in DL) types of architectures in DNs, as they can both reduce the need for training.

## Discussion

The lack of interpretability in diffractive optical networks for high-dimensional tasks has been a longstanding challenge. The difficulty of interpretation arises mainly in two aspects. First, the iterative algorithms used for training typically generate seemly random masks that are not intuitive at first glance. Second, there is a large number of pixels and layers with unknown relations. In this work, we have shown that, for a high-dimensional vortex mode sorting task, the trained diffractive optical networks become physically interpretable as the mathematical operations of each diffractive surface layer have been unveiled after the training. This is done by inspecting the propagating and diffracted field distribution in each layer and extracting the collective effects within the system. This led to the first physically interpretable diffractive optical networks, as demonstrated here for the successive diffractive layers. To our knowledge, there have been no attempts to interpret the physics behind each layer of the complex multi-layer diffractive structure for mode sorting tasks, even though the system is known to perform linear transformations. We demonstrate that such interpretation can be achieved, even when the DN is designed by iterative algorithms.

We started with a simple vortex mode sorting example and showed how the conventional, less interpretable DN can be turned into an interpretable diffractive network by forwardly propagating the input modes and analyzing the intermediate mode states. Surprisingly, our two-layer DN reproduces the famous log-polar transformation and focusing procedure for vortex mode sorting. Next, we investigate the collective physical phenomena behind multiple diffractive layers for the high-dimensional LG mode sorting task. First, we demonstrated that the increase in classification dimensionality can be incorporated with an additional network layer, while maintaining the independence of both physical processes. While previous research mainly focused on information capacity limits[36], we offer a distinct perspective by examining the number and evolution of transformations within a DN with varying layer counts. We have also discovered an intriguing physical transformation division phenomenon when the diffractive layer of the network increases, and observed redundant physical solutions in DNs when two sequential diffractive layers work synergistically to achieve only one transformation, thereby enhancing our understanding of collective effects within the system. The interesting underfitting/overfitting effect for physical transformation was accompanied by the increased/decreased performance of trained DNs, which can be further utilized for designing DNs with appropriate mask numbers. Finally, based on the understanding of the transformation processes behind the networks, we demonstrated how such a physical interpretation can be used to extrapolate the trained network under physical perturbations, thereby eliminating the need for re-training or realignment.

In this work, we have specifically focused on the LG mode sorting process, which is a well-known problem in the field of optics. Essentially, any two-dimensional complex input field can be decomposed into a linear superposition of any orthogonal modes that form a complete set[37]. Thus, while this work is only devoted to vortex mode sorting, the insights gained from interpreting the diffractive layers could be crucial for a variety of optical tasks, including image classification and beyond[38, 39]. One of the examples is the recently reported image denoising processor, which can be viewed as a spatial filtering system. The symmetry of the mask distribution with the frequency filter at the middle layer is observed in that work (see the supporting information of the original literature for further discussion)[40]. Finally, while the concept demonstrated here is for linear diffractive networks only, it has recently been reported that it is possible to perform nonlinear computing on similar network structures[41, 42]. This work will inspire further research on deciphering the mathematical meaning of complex optical networks with nonlinearities[43-46], and eventually on other deep neural networks in DL, where the black box problem is mainly caused by increased system depth and complex nonlinearity[47, 48].

## Materials and Methods

### Model training and numerical calculation

The diffractive networks are trained by adapting the widely used wavefront matching approach[15]. In all trainings, each mask layer is defined as a complex phase matrix with initial values of unitary (flat surfaces). The angular spectrum method is applied to compute the free-space propagation of the field over varying distances. The gradient-descent optimization procedure is applied, with the target function defined as output mode correlation coefficients. We reduce the degree of freedom of the network by enforcing the phase smoothness through k-space filtering (with a numerical aperture of 0.58)[49, 50]. The masks are 640 × 640 in size, with a pixel size of 8 μm × 8 μm. The wavelength of the input modes is 1550 nm. Other system parameters depend on the designed task and are listed in the following.

For the case of the vortex mode sorter with two masks (Fig. 2), we used LG modes with a beam waist parameter of 400 μm. The interplane distance is set to be 20 mm, while the observation plane is 100 mm away from the last network surface. As mentioned in the main text, our dataset consists of 11 LG modes (with p=0) with l from -5 to 5, while l = 4, 2, and -2 are included in the test set, and the rest are used as the training dataset. We used resultant fields from focused rectangle shapes of different phase gradients as the target output fields to train our DN. The same trained masks are used in Fig. 5a to extrapolate the masks for different combinations of wavelengths and/or distances. The detection matrix is obtained following the approach in typical mode sorting work[12].

For the case of the LG mode sorter with three masks (Fig. 3), LG modes with a beam waist parameter of 170 μm are given as inputs. The interplane distances are 26.8 mm, while the observation plane is 50 mm away from the last network surface. Target resultant fields are Gaussian beams at the waist with a value of 55 μm. The center-to-center distances for neighbor modes are set to be 100 μm. Our dataset consists of 33 $LG_{l,p}$ modes with l from -5 to 5 and p from 0 to 2, while the cases $LG_{-2,2}$, $LG_{3,1}$, $LG_{-3,1}$, $LG_{4,0}$, $LG_{1,0}$ are included in the test set, and the rest are used as the training dataset. We note that the choice of training set and test set does not have observable effects on the results, which is due to the symmetries of the mode index spectrum[23]. Similar settings are applied for the LG mode sorter shown in Fig. 4, except for the total number of diffractive layers. The same trained masks are used in Fig. 5b to extrapolate the masks for different combinations of wavelengths and/or distances.

There are a few other points that need to be discussed in our numerical results. Under the initial condition of flat surfaces, the training process is rather stable, as it always leads to physically identical solutions (Fig. S1). However, if the initial surface states are assigned with fully random phases (phase uniformly distributed from 0 to $2\pi$), one can obtain trained networks with different mask distributions (Fig. S2a). Interestingly, the systems have performance comparable to the ones trained with initially flat surfaces (Fig. S2b). While phase smoothness can still be observed in the trained masks, the study of their physical meaning is challenging and seemly unnecessary, as these "randomly trained" masks vary for different realizations (Fig. S2c). What is intriguing is, the solutions become stable and repeatable when the randomness is reduced, which leads to trained results identical to the flat surface conditions (Fig. S2d). This implies that the trained DNs are more robust against the choice of initial parameters, compared to digital neural networks.

**Experimental setups**

In our experiment, we used two cascaded multi-plane light conversion (MPLC) systems. Principles and illustrations of the MPLC system for the generation and demultiplexing of modes can be found in other literature[14, 51]. Briefly, each MPLC system contains one spatial light modulator (Holoeye PLUTO-2.1, pixel number: 1920 × 1080, pixel size: 8 μm × 8 μm), together with a reflective mirror. The light source (Keysight N7714A, wavelength: 1550 nm) we used has a beam waist of 110 μm. The first MPLC generates high-quality LG modes with tunable beam waists and mode indices. For each LG mode, three sequential masks are trained and applied with the same protocol. The trained or numerical masks are applied in the second MPLC system and perform the sorting task. The detector (Hamamatsu, C12741-03, pixel number: 640 × 512, pixel size: 20 μm × 20 μm) is placed at the output plane to capture the sorted mode intensity. The incident angle of the MPLC system is set to be 11 degrees, which minimizes the mode distortion.